\documentstyle[aps,prl]{revtex} 
\tighten
\begin{document}
\input epsf.sty
\flushbottom
\draft
\twocolumn[\hsize\textwidth\columnwidth\hsize\csname @twocolumnfalse\endcsname 
\title{
On the Nature of Charge Transport in Quantum-Cascade Lasers 
}
\author{Rita Claudia Iotti$^{1,2}$ and Fausto Rossi$^{1}$}
\address{
$^{1}$Istituto Nazionale per la Fisica della Materia (INFM) and 
Dipartimento di Fisica,\\ 
Politecnico di Torino, Corso Duca degli Abruzzi 24, 10129 Torino, Italy \\
$^{2}$Scuola Normale Superiore, Piazza dei Cavalieri 7, 56126 Pisa, Italy
}

\date{\today}
\maketitle

\begin{abstract}

The first {\it global quantum simulation} of semiconductor-based 
quantum-cascade lasers is presented. 
Our three-dimensional approach allows to study in a purely microscopic 
way the current-voltage characteristics of state-of-the-art unipolar 
nanostructures, and therefore to answer the long-standing controversial 
question: {\it is charge transport in quantum-cascade lasers mainly 
coherent or incoherent?} 
Our analysis shows that: 
(i) Quantum corrections to the semiclassical scenario are minor; 
(ii) Inclusion of carrier-phonon and carrier-carrier scattering gives 
excellent agreement with experimental results.

\end{abstract}

]

\narrowtext

Since the seminal paper of Esaki and Tsu~\cite{Esaki}, semiconductor-based 
nanometric heterostructures have been the subject of an impressive 
theoretical and experimental activity, due to their high potential 
impact in both fundamental and applied research~\cite{B-R,T-A}.
One of the main fields of research focuses on exploiting ``band-gap 
engineering'', namely the splitting of the bulk conduction band into 
several subbands, to generate and detect 
electromagnetic radiation in the infrared spectral region, as originally 
envisioned by Kazarinov and Suris~\cite{KS}.

Unipolar coherent-light sources like quantum-cascade lasers 
(QCLs) \cite{QCL94} 
are complex devices, whose core is a multi-quantum-well (MQW) 
structure made up of repeated stages of active regions sandwiched between
electron-injecting and collecting regions. When a proper bias is applied,
an ``electron cascade'' along the subsequent quantized-level energy staircase 
takes place. 
QCLs are usually modelled in terms of {\it n}-level 
systems~\cite{QCL}. 
%Their theoretical description is thus often grounded on purely macroscopic 
%models, which neglect the existence of transverse or ``in-plane'' degrees 
%of freedom. 
As pointed out in~\cite{APL}, such a macroscopic modeling can only operate 
as an {\it a posteriori} fitting procedure.
In contrast, 
for a detailed understanding of the basic physical processes involved,
a fully three-dimensional (3D) description is needed.
More specifically, two main issues need to be addressed: 
(i) the nature of the hot-carrier relaxation within the device active region; 
(ii) the nature ---coherent versus incoherent--- of the 
physical mechanisms governing charge transport through 
injector/active-region/collector interfaces.

Point (i) has been recently addressed in \cite{APL}, where the usual 
macroscopic treatment of the device active region has been compared to a 
fully kinetic description, based on a Monte Carlo (MC) solution~\cite{MC} of 
the following set of equations:
\begin{equation}
{d \over dt} f_{{\bf k}\nu} =
\left[
g_{{\bf k}\nu} - \Gamma_{{\bf k}\nu} f_{{\bf k}\nu}
\right]_{{\rm i/l}} 
+ \! \sum_{{\bf k}'\nu'} \! \left[
P_{{\bf k}\nu,{\bf k}'\nu'} f_{{\bf k}'\nu'} 
- P_{{\bf k}'\nu',{\bf k}\nu} f_{{\bf k}\nu} 
\right]\ . 
\label{PPE}
\end{equation}
Here, the first two terms describe ---still on a partially phenomenological
level--- injection/loss (i/l) of carriers with parallel or in-plane 
wavevector ${\bf k}$ in 
subband $\nu$, while the last ones describe intra- as well as inter-subband 
in- and out-scattering processes (${\bf k}\nu \to {\bf k}'\nu'$).
%within the device active region only.
As reported in \cite{APL}, the quantum-cascade within the active region is 
mainly governed by LO-phonon emission. 
%\cite{active-region}.
However, such a microscopic analysis, being limited to the device 
active region only, does not allow to answer point (ii); 
%which are the basic physical mechanisms governing charge transport through 
%injector/active-region/collector interfaces?
This issue is intimately related to the long-standing controversial 
question~\cite{RT}: 
{\it is charge transport in quantum-cascade lasers mainly coherent or 
incoherent?}

To provide a definite answer to this fundamental question, 
we present the first {\it global quantum simulation} 
---injector plus active region plus collector--- 
of semiconductor-based QCL structures.
To this end, two basic steps are needed:
First, the partially phenomenologic model in (\ref{PPE}) has to be 
replaced by a fully microscopic description of the whole MQW core 
structure; Second, the semiclassical or Boltzmann-like treatment will be 
replaced by a fully quantum-mechanical one.
These are not trivial tasks. 

The first step requires 
%in addition to a proper description of all the 3D electron states within the 
%active as well as injector/collector regions \cite{APL}, 
a proper simulation scheme to ``close the circuit'' without resorting to
phenomenological i/l parameters. 
To this aim, given the set of 3D single-particle electron 
states $\{{\bf k}\nu\}$ corresponding to a single QCL stage,
we consider the ideal MQW structure obtained 
as infinite repetition of this QCL periodicity region (see Fig.~\ref{fig1}).
Within such extended scheme, the time evolution of the carrier distribution 
function $f_{{\bf k}\alpha}$ is governed by the following Boltzmann-like 
equation:
\begin{equation}
{d \over dt} f_{{\bf k}\alpha} =
\sum_{{\bf k}'\alpha'} \! 
\left[
P_{{\bf k}\alpha,{\bf k}'\alpha'} f_{{\bf k}'\alpha'} 
- P_{{\bf k}'\alpha',{\bf k}\alpha} f_{{\bf k}\alpha} 
\right] \ .
\label{BTE}
\end{equation}
Here, $\alpha \equiv (\lambda,\nu)$ denotes the generic electron state in 
our MQW structure, i.e., the $\nu$-th state of the $\lambda$-th stage.
To ``close the circuit'', we impose periodic boundary conditions limiting 
the inter-stage ($\lambda' \ne \lambda$) scattering to just nearest-neighbor 
coupling ($\lambda' = \lambda \pm 1$).
In view of the translational symmetry, we are allowed to simulate carrier 
transport over the central ---i.e., $\lambda = 0$--- stage 
only~\cite{periodicity}.

We have applied the above ---still semiclassical--- global-simulation scheme 
to state-of-the-art QCL structures.
As prototypical device, we have considered the 
GaAs/(Al,Ga)As-based diagonal-configuration QCL in \cite{Sirtori98}, 
schematically depicted in Fig.~\ref{fig1}, in which our simulation strategy 
is also sketched. 
Here, the energy levels and probability densities of various electron states 
within the simulated stage ($\lambda = 0$) are plotted: 
They correspond to the device active region ($\nu = 1,2,3$ according to the 
standard notation) as well as to the collector region ($\nu =$ A,B,C,D,E). 
%The nearest stages $\lambda = \pm 1$ are partially shown for clarity as 
%well.
%
In order to properly model phase-breaking hopping processes, in 
addition to carrier-optical phonon scattering, all various intra- as 
well as intersubband carrier-carrier interaction mechanisms have been 
considered \cite{other-scat-mech}. 
%To this end, the well-established MC method for the simulation of 
%intercarrier scattering in quantum-well systems~\cite{MC} has been 
%employed.

As a starting point, we have investigated the relative weigth of the 
carrier-carrier and carrier-phonon competing energy-relaxation channels.
The time evolution of the carrier population in the various subbands as 
well as of the total current density 
%in the presence of carrier-phonon scattering 
are depicted in Fig.~2.
Parts (a) and (b) report the population dynamics without and with 
two-body carrier-carrier scattering, respectively.
In our ``charge-conserving'' scheme, we start the simulation assuming the 
total number of carriers to be equally distributed among the different 
subbands; then the electron distribution functions evolve according to 
Eq. (\ref{BTE}) and a steady-state condition is eventually reached, 
leading to the desired $3 \rightarrow 2$ population inversion.
As we can see, the inclusion of intercarrier scattering has significant 
effects:
It strongly increases inter-subband carrier redistribution, thus reducing 
the electron accumulation in the lowest energy level A and 
optimizing the coupling between active region and injector/collector 
(the populations of subbands 3 and B get equal).
This effect comes out to be crucial in determining the electron flux 
through the MQW structure. 
Figure~2(c) shows the simulated current-voltage characteristics of our QCL 
device, obtained with and without carrier-carrier interaction.
At the threshold operating parameters estimated in~\cite{Sirtori98} 
[marked by an arrow in Fig.~2(c)] the current density in the presence of 
both electron-phonon and electron-electron scattering mechanisms is about 
4~kA/cm$^{2}$. This value is in relatively good agreement with 
experiments~\cite{agreement}.

The results plotted in fig.~\ref{fig2} clearly demonstrate that 
within a purely semiclassical picture the 
electron-phonon interaction alone is not able to efficiently couple the 
injector subbands to the active region ones: 
While carrier-phonon relaxation well describes 
the electronic quantum cascade within the bare active region~\cite{APL}, 
carrier-carrier scattering plays an essential role in determining 
charge transport through the full core region.
This can be ascribed to two typical features of carrier-carrier interaction 
---compared to the case of carrier-phonon---:  
(i) this is a long-range two-body interaction mechanism, which also couples 
non-overlapping single-particle states [see Fig.~1];
(ii) the corresponding scattering process at 
relatively low carrier density is quasielastic, thus coupling nearly 
resonant energy levels, like states 3 and B.

So far, 
%due to our semiclassical description, 
no quantum mechanical effects, like coherent resonant tunneling between 
adjacent states, have been considered. 
In order to see how such coherent phenomena can change the scenario presented 
so far, we have extended our semiclassical simulation scheme in terms of a 
density-matrix formalism \cite{DMT}.
%, which allows to describe on the same footing phase 
%coherence as well as energy relaxation and dephasing 
In the proposed quantum-transport approach the basic ingredient is the 
single-particle density matrix 
$\rho_{ij} = 
\left\langle a^{\dagger}_{i} a^{ }_{j} \right\rangle$,
where $a^{\dagger}_i$ ($a^{ }_i$) denote creation (destruction)
operators for a carrier in state $i \equiv {\bf k}\alpha$ \cite{DM}.
%Within the usual mean-field approximation and Markov limit the 
Its time evolution 
%of the density matrix 
is given by:
\begin{equation}\label{SBE}
\frac{d}{dt} \rho_{ij} = -i \omega_{ij} \rho_{ij} +
\sum_{i'j'} \left[ \left(
\Gamma^{{\rm in}}_{ij,i'j'} \rho_{i'j'} -
\Gamma^{{\rm out}}_{ij,i'j'} \rho_{i'j'} 
\right) 
+ \mbox{c.c.} \right] \ ,
\end{equation}
where $\hbar\omega_{ij} = \epsilon_i - \epsilon_j$ is the energy 
difference between states $i$ and $j$.
Here, the first term describes the coherent evolution of the noninteracting 
carrier system while the second contribution describes energy relaxation as 
well as dephasing, 
in terms of the generalized in- and out-scattering superoperators $\Gamma$ 
\cite{Gamma}.
Equation (\ref{SBE}) is the desired quantum-mechanical generalization of the 
Boltzmann transport equation in (\ref{BTE}). 
%indeed, by neglecting all 
%non-diagonal terms of the single-particle density matrix
%($\rho_{ij} = f_i \delta_{ij}$), the latter is 
%easily recovered \cite{Gamma}.

Analogous to our semiclassical simulation scheme, for the new 
quantum-transport formalism we can adopt the same periodic conditions to 
``close the circuit''.
%This allows us to study the density matrix evolution in the $\lambda =0$ 
%periodicity region only, accounting for diagonal as well as non-diagonal 
%scattering processes within region $\lambda = 0$ as well as from and to 
%regions $\lambda = \pm1$. 
%this again will provide the current-voltage 
%characteristics without resorting to any phenomenological parameter.
Moreover, since for the QCL design considered in-plane and along-{\it z}
carrier dynamics are strongly decoupled \cite{fact},
it is possible to adopt a factorization of the density matrix according to
$\rho_{ii'} = \rho_{{\bf k}\nu,{\bf k}'\nu'} = \rho_{\nu\nu'} 
f^\parallel_{\bf k} \delta_{{\bf k, k'}}$,
where $f^\parallel$ denotes the parallel or in-plane carrier distribution. 
%\cite{fact}.
This allows us to obtain an effective equation of motion for 
$\rho_{\nu\nu'}$: the latter, which has again the structure of Eq.~(\ref{SBE}), 
involves effective scattering matrices, given by an in-plane average of the 
quantities $\Gamma^{\rm in/out}$ in (\ref{SBE}).

Comparing the results obtained with the proposed quantum-transport 
approach to those of the semiclassical global simulation scheme, we find 
negligible quantum corrections (of a few percents) to the stationary current 
density. In the absence of carrier-carrier scattering, we get, e.g., a 2\% 
quantum correction to the result at threshold reported in Fig.~2. 
This is due to the extremely small value of non diagonal density-matrix 
elements $\rho_{\nu \ne \nu'}$ (compared to the diagonal ones).
The physical interpretation of such a behaviour proceeds as follows: 
Non-diagonal elements in
%the generalized carrier-phonon scattering operators 
$\Gamma^{\rm in/out}$ tend to maintain a non-diagonal density matrix also 
in stationary conditions. On the other hand, diagonal energy-relaxation and 
dephasing processes tend to suppress non-diagonal terms on the sub-picosecond 
time-scale. 
Since the average device transit time 
%---i.e., the time needed by an electron to 
%travel through the device period and thus to be reinjected into the simulation 
%region--- 
is of the order of several picoseconds, the degree of coherence, i.e., 
the weight of non-diagonal density-matrix elements, in stationary conditions 
is very small.
This does not mean that coherent phenomena, like resonant-tunneling processes, 
are not present.
Figure 3 presents the first simulation of the ultrafast dynamics of a 
properly tailored electron wavepacket, within the MQW core region of the QCL
sketched in Fig.~1. The aim is to focus on the injector---active-region 
tunneling mechanisms. For this reason, the system has been prepared
at time $t$ = 0 to reproduce a charge-density distribution fully localized in 
the $\lambda$ = 0 injector and compatible with resonant tunneling into the 
next-stage active region (see (0,B)-(+1,3) alignment in Fig.~1).
As clearly shown, the transient dynamics is characterized by a strong interplay 
between phase-coherence and relaxation (on a picosecond time-scale); Only 
at much longer times it will eventually reach the stationary transport 
solution, in which incoherent sequential tunneling is the dominant interwell 
mechanism~\cite{ultrafast}.

In summary, we are in the position to answer the long-standing controversial 
question on 
the nature ---coherent versus incoherent--- of charge transport in QCLs 
previously mentioned: 
For the typical structures considered, energy-relaxation and dephasing 
processes are so strong to destroy any phase-coherence effect on a 
sub-picosecond time-scale; as a result, the usual semiclassical or incoherent 
description of stationary charge transport is found to be in excellent 
agreement with experiments.
%
%In summary, we have presented the first global quantum simulation 
%---steady-state as well as ultrafast--- of semiconductor-based QCLs. 
%The proposed theoretical scheme allows for a fully microscopic evaluation of 
%the current-voltage characteristics of prototypical QCL structures. 
%The excellent agreement between our semiclassical and quantum simulations 
%and recent experimental results confirms the incoherent nature of charge 
%transport in such unipolar quantum devices.

\medskip\par
%Discussions with S.~Barbieri, F.~Beltram, C.~Sirtori and A.~Tredicucci are 
%gratefully acknowledged. 
We are grateful to F. Beltram, C. Sirtori and A. Tredicucci for fruitful conversations 
and discussions. This work has been partially supported by the European Commission through 
the Brite Euram {\sc Unisel} project and the {\sc Fet Wanted} project.
\begin{figure}
%\hspace*{0cm} \epsfxsize7cm \epsfbox{f1.eps} 
%\vspace*{0.5cm}
\caption{\label{fig1} 
Schematic representation of the conduction-band profile along the growth
direction for the diagonal-configuration QCL structure of 
Ref.~\protect\cite{Sirtori98}. 
The MQW is biased by an electric field of 48 kV/cm.
The levels $\nu = 1,2,3$ and $\nu =$ A,B,C,D,E in the active and 
collector regions (full and dashed lines, respectively) of the simulated 
stage ($\lambda = 0$) are also plotted together with the corresponding 
probability densities. 
The replica of level 3 in the following stage $\lambda = +1$ is shown for 
clarity.
}
\end{figure}
\begin{figure}
\caption{\label{fig2} 
Time evolution of simulated carrier densities in the various subbands of 
the MQW structure of Fig.~\protect\ref{fig1} (full lines: $\nu =$ A,B,C,D,E; 
dotted lines: $\nu =$ 1,2,3), without (a) and with (b) intercarrier scattering. 
(c): simulated applied-field vs current-density characteristics of the
whole structure at 77 K, in presence (discs) and absence (filled squares) of
carrier-carrier interaction. Threshold applied field (48 kV/cm) is marked by 
an arrow. Dashed lines are a guide to the eye.
%An electric field of 48 kV/cm has been applied.
}
\end{figure}
\begin{figure}
\caption{\label{fig3} 
Time evolution of the charge density for an electron wavepacket properly 
tailored to study the carrier tunneling dynamics accross the injection barrier,
for the QCL design of Fig.~1.
At $t$ = 0 ps (lower panel (c)) the wavepacket is fully localized in the 
injector. Shaded regions correspond to (Al,Ga)As barriers in the heterostructure 
design.
The transient dynamics (middle panel (b)) is characterized by a strong interplay 
between phase-coherence and relaxation processes.
At much longer times (upper panel (a)) the system will eventually evolve into the 
stationary transport solution.
}
\end{figure}

\begin{references}
\bibitem{Esaki} L.~Esaki and R.~Tsu, IBM J. Res. Develop., {\bf 14},
61 (1970).

\bibitem{B-R}
See, e.g., 
J. Shah, {\it Ultrafast Spectroscopy of Semiconductors and Semiconductor
Nanostructures} (Springer, Berlin, 1996);
{\it Theory of Transport Properties of
  Semiconductor Nanostructures}, edited by E. Sch\"oll
(Chapman \& Hall, London, 1998).

\bibitem{T-A}
See, e.g., 
{\it Physics of Quantum Electron Devices}, edited by F. Capasso
(Springer, Berlin, 1990).

\bibitem{KS} 
R.~F.~Kazarinov and R.~A.~Suris, Fiz. Tekh. Poluprov. {\bf 5}, 797
(1971); Sov. Phys. Semicond. {\bf 5}, 707 (1971).

\bibitem{QCL94}
J.~Faist {\it et al.}, Science {\bf 264}, 553 (1994);
G.~Scamarcio {\it et al.}, Science {\bf 276}, 773 (1997);
%F.~Capasso {\it et al.}, Phys. World {\bf 12}, 27, (1999);
C.~Gmachl {\it et al.}, Science {\bf 286}, 749 (1999). 

\bibitem{QCL}
The theoretical description of QCLs is often grounded on purely macroscopic 
models, which neglect the existence of transverse or ``in-plane'' degrees
of freedom.
%The basic idea is to identify a three-level quantum structure: 
%carriers are injected into the upper level (level 3) which is 
%``metastable'', i.e., weakly coupled to the lower levels; in contrast, 
%carriers in the intermediate level (level 2) can relax very fast into the 
%lowest level (level 1).
%As a result, the desired population inversion between levels 3 and 2
%is obtained. 

\bibitem{APL}
R. C. Iotti and F. Rossi, Appl. Phys. Lett. {\bf 76}, 2265 (2000), and references 
therein.

\bibitem{MC}
See, e.g., C. Jacoboni and P. Lugli, {\it The Monte Carlo Method for 
Semiconductor Device Simulations} (Springer, Vienna, 1989).
%S.M. Goodnick and P. Lugli, in {\it Hot Carriers in Semiconductor
%Nanostructures: Physics and Applications}, edited by J. Shah (Academic,
%San Diego, 1992), p. 191.

%\bibitem{active-region}
%Due to the relatively large intersubband energy splitting, coherent phenomena
%---not included in \cite{APL}--- are expected to play no significant role; 
%this is confirmed by the quantum-transport simulation of Fig.~3.

\bibitem{RT} C. Sirtori {\it et al.} IEEE J. Quantum Electron.
{\bf 34}, 1722 (1998).

%\bibitem{co-inco}
%With coherent vs. incoherent transport, here we mean the degree 
%of quantum-mechanical phase correlation between two generic single-particle
%states in the structure.
%Within a strongly simplified picture, it is usually stated that the charge
%in the collector region comes to a quasiequilibrium and, acting
%as an incoherent source, 
%is then injected into the active lasing region
%(incoherent sequential tunneling). In contrast, 
%a proper treatment of QCLs requires a 
%global description of injector + active region + collector, which may imply
%the presence of a residual phase coherence between injector and 
%active region (coherent resonant tunneling).

%\bibitem{Barbieri99}
%S. Barbieri, F. Beltram, and F. Rossi, Phys. Rev. B {\bf 60}, 1953 (1999).
%%and references therein.

\bibitem{periodicity}
Every time a carrier in state $\nu$ undergoes an inter-stage scattering 
process (i.e., $0,\nu \to \pm 1,\nu'$), it is properly reinjected 
into the central region ($0,\nu \to 0,\nu'$) and the corresponding 
electron charge $\pm e$ will contribute to the current through the 
device. This allows for a purely microscopic evaluation of the 
current-voltage characteristics.
%We stress that, contrary to Eq.~(\ref{PPE}), the proposed global-simulation 
%scheme allows to study in a purely microscopic way ---without resorting to 
%phenomenological i/l parameters--- the current-voltage characteristics of 
%state-of-the-art unipolar nanostructures.

\bibitem{Sirtori98}
C. Sirtori {\em et al.}, Appl. Phys. Lett. {\bf 73}, 3486 (1998).

\bibitem{other-scat-mech}
Other scattering mechanisms not included in the simulation, e.g., 
carrier---acoustic-phonon coupling, are expected to play a minor role~\cite{MC}. 
%In particular, the interaction with acoustic phonons, inspite of its 
%quasi-elastic nature, does not affect charge transport significantly due to
%its small coupling constant. 

\bibitem{agreement}
The apparent discrepancy between the theoretical and the experimental
($\simeq$ 7 kA/cm$^2$) results is due, we believe, to the different estimate 
of the potential drop per period required to line up the ground state A of 
the injector with level 3 of the active region.
Indeed, our Schr\"odinger-Poisson calculation predicts a good alignment 
at a higher bias, which agrees much better with the measured 
one~\protect\cite{Sirtori98}. 
%As shown in Fig.~\protect\ref{fig2}(c) the 
%simulated current density at
%68 kV/cm is about 7 kA/cm$^2$, thus enforcing the present interpretation.

\bibitem{DMT}
T.~Kuhn, in {\it Theory of Transport Properties of Semiconductor 
Nanostructures}, edited by E. Sch\"oll (Chapman \& Hall, London, 1998),
p. 173.
%F. Rossi, A. Di Carlo, and P. Lugli, Phys. Rev. Lett. {\bf 80}, 3348 (1998).

\bibitem{DM}
This is defined as the average value of two creation and destruction 
operators:
its diagonal elements correspond to the usual distribution functions 
$f_{{\bf k}\alpha}$ of the semiclassical Boltzmann theory 
while the off-diagonal terms ($i \ne j$) describe the degree of 
quantum-mechanical phase coherence between states $i$ and $j$.

\bibitem{Gamma}
The in- and out-scattering matrices $\Gamma$ in (\ref{SBE}) describe the effect 
on the time evolution of the density-matrix element $\rho_{ij}$ due to the 
generic element $\rho_{i'j'}$.
Their diagonal parts, i.e., $ii' = jj'$, correspond to the semiclassical 
scattering rates $P_{ii'} \equiv P_{{\bf k}\alpha,{\bf k}'\alpha'}$ 
in Eq.~(\ref{BTE})~\protect\cite{DMT}. 

\bibitem{fact}
R.C. Iotti and F. Rossi, Appl. Phys. Lett. {\bf 78}, 2902 (2001).
%Such factorized form is well justified by the strong in-plane carrier 
%thermalization reported in [R.C. Iotti and F. Rossi, Appl. Phys. Lett. 
%{\bf 78}, 2902 (2001)]. This confirms the strong separation between in-plane and 
%$z$ degrees of freedom.
%Moreover, in view of the homogeneous in-plane charge distribution, 
%non-diagonal density-matrix elements (${\bf k} \ne {\bf k'}$) are equal to 
%zero.

\bibitem{ultrafast}
These results suggest that ultrafast optical experiments, like pump-and-probe 
or four-wave-mixing measurements, should provide a clear fingerprint of such 
a coherent vs. energy-relaxation carrier dynamics.
This is confirmed by recent ultrafast experiments by Eickemeyer {\it et 
al.} [F.~Eickemeyer {\it et al.}, to appear in the Proceedings of the 
12$^{\rm th}$ International Conference on Nonequilibrium Carrier Dynamics 
in Semiconductors (HCIS12, Santa Fe, 2001)].

\end{references}
\end{document}